\documentstyle[preprint,aps]{revtex}
\begin{document}
\title{A class of two-dimensional Yang--Mills vacua and their
relation to the non-linear sigma model}
\author{A. Bassetto ($^*$)}
\address{CERN, Theory Division, CH-1211 Geneva 23, Switzerland\\
INFN, Sezione di Padova, Padua, Italy}
\author{G. Nardelli}
\address{Dipartimento di Fisica, Universit\`a di Trento,
38050 Povo (Trento), Italy \\ INFN, Gruppo Collegato di Trento, Italy}
\maketitle
\begin{abstract}
Classical vacuum - pure gauge -  solutions of Euclidean two-dimensional $SU(2)$
Yang--Mills theories are studied. Topologically non-trivial vacua are found
in a class of gauge group elements isomorphic to $S_2$.
These solutions  are unexpectedly  related to the solution of the non-linear
$O(3)$ model and to the motion of a particle in a
periodic potential.
\end{abstract}


\vskip 5.0truecm

\noindent
($^*$) On leave of absence from
Dipartimento di Fisica, Via Marzolo 8 -- I-35131
Padova.
\vfill\eject

\narrowtext
A Coulomb gauge description of Yang--Mills fields in three spatial dimensions is
ambiguous\cite{gri}: several  gauge-equivalent potentials $A_\mu$ represent the
same physical field configuration, all these potentials satisfying the Coulomb
gauge and  being related by {\it large} gauge transformations, {\it i.e.}
gauge transformations that cannot be smoothly deformed to the identity.
Consequently, although gauge-equivalent, these field configurations are
topologically inequivalent. Clearly, this ambiguity persists
in the pure-gauge solutions, leading to different vacua \cite{jac1}.

Topological arguments suggest that this pathology disappears in
two spatial dimensions. Nevertheless, the situation changes when
considering
gauge group elements belonging to a particular class of $SU(2)$.
Topologically inequivalent vacua can indeed be found,
which are related to the solutions of the non-linear $O(3)$ model
on a plane and to the solutions
of a particle moving in a periodic
potential.

The purpose
of this note is to illustrate this phenomenon.

Our conventions are as follows. Space coordinates will be
denoted by $x_{\mu}$, $\mu=1,2$; $T^{j}=\sigma^j/2i$, $j=1,2,3$, will denote
the anti-Hermitian $SU(2)$ generators; the first two components of the internal
indices will be denoted by $a=1,2$.

Classical vacuum solutions are pure gauges,  $A_\mu = (1/g) U^{-1}
\partial_\mu U$, and we shall look for non-trivial vacuum solutions satisfying
the  gauge choice $\partial_\mu A_\mu =0$, just as in the standard Gribov
copies in three spatial dimensions. The $SU(2)$ gauge
group element $U$ will be chosen to belong to the class of transformations
not depending on $\sigma^3$, namely
\begin{equation}
U= n_0 - i \sigma^a n^a\ .
\label{u3}
\end{equation}
These elements in general are {\it not} a subgroup of $SU(2)$.
Rather, since $n_0^2 + n^a n^a =1$,
they belong to a class isomorphic to $S_2$, which is in turn a
homogeneous space of $SU(2)$.
The pure  gauge potential associated to (\ref{u3}) is
\begin{equation}
gA_\mu =-2 T^3 \epsilon^{ab} n^a\partial_\mu n^b + 2 T^a (n_0\partial_\mu n^a -
n^a\partial_\mu n_0)
\label{a3}
\end{equation}
and the gauge condition $\partial_\mu A_\mu=0$ implies the following three
equations
\begin{equation}
\epsilon^{ab} n^a \Delta n^b =0 \ \ , \qquad \ \  n^a\Delta n_0 - n_0 \Delta
n^a = 0\ \ .
\label{n1}
\end{equation}
In turn, eqs. (\ref{n1}) can be rewritten in a more compact form by renaming
$n_0\equiv n^3$, so that the requirement   $U\in SU(2)$
in terms of $n^i$  becomes $n^i n^i = 1$, $i=1,2,3$, and ${\bf
n}=\{n^i\}$
satisfies the same constraint as a non-linear $O(3)$ model. Then
eqs. (\ref{n1}) simply become
\begin{equation}
\epsilon^{ijk} n^j \Delta n^k =0\ \ .
\label{n2}
\end{equation}
 Any field configuration $n^i$ satisfying the $O(3)$
self-duality (or anti-self-duality) conditions
\begin{equation}
\partial_\mu n^i = \pm \epsilon^{ijk} \epsilon_{\mu \nu}\partial_\nu n^j n^k
\label{sd}
\end{equation}
is a solution of eqs. (\ref{n2}), as can be easily checked by direct
inspection. It is important to stress  that eqs. (\ref{n2}) are {\it not} the
equations of motion of the non-linear $O(3)$ model [in the non-linear $O(3)$
model the fields $n^i$ satisfy the equations $\Delta n^i = n^i n^j \Delta
n^j$]. Nevertheless, the same self-duality condition that solves the non-linear
$O(3)$ model also solves  eqs.  (\ref{n2}).

Any  field configuration satisfying the constraint $n^in^i=1$
has a topological
invariant defined by
\begin{equation}
Q={1\over 8\pi} \int d^2 x \, \epsilon^{ijk} \epsilon_{\mu \nu}n^i \partial_\mu
n^j \partial_\nu n^k \ \in {\bf Z}\ ,
\label{top}
\end{equation}
which is obviously the topological invariant of the non-linear $O(3)$ model as
well. The integer number $Q$ simply tells how many times the stereographic
projection of the Euclidean plane wraps onto the sphere $n^in^i=1$ in the change
of variables $\{x^\mu\}\to \{ n^i\}$.
Self-dual configurations (upper sign in eq. (\ref{sd})) have positive $Q$,
whereas for anti-self-dual fields (lower sign in (\ref{sd})) $Q<0$. Since the
two classes of solutions are  related by a parity transformation,
it is not restrictive to consider only   one  sign in eq.
(\ref{sd}). For definiteness, from now on we shall discuss self-dual
configurations.

 The way of solving eq. (\ref{sd}) is standard
\cite{pol}: perform a stereographic projection of the sphere $n^i n^i=1$ onto
the plane $n^1\times n^2$ and consider the stereographically projected variables
$\omega^1 = n^1/(1-n^3)$, $\omega^2 = n^2/(1-n^3)$. Then, eq.
(\ref{sd}), written in terms of the function $\omega = \omega^1 + i
\omega^2$,
simply becomes the Cauchy--Riemann conditions, so that any meromorphic
function $\omega$ is a solution. Once a meromorphic function
$\omega(z)$ is chosen, the $n^i$ can easily be obtained
by inverse stereographic projection
\begin{equation}
n_0= n^3 = {|\omega(z)|^2 -1 \over |\omega(z)|^2 +1}\ \ \ \ ,\qquad \ \ \  n^a=
{2\omega^a \over |\omega(z)|^2 +1}\ .\ \
\label{n3}
\end{equation}
The topological number $Q$ can be written in terms of $\omega$ as
\begin{equation}
Q={1\over \pi} \int d^2x\, {|\omega'(z)|^2\over (1 + |\omega(z)|^2)^2}\ .
\label{top2}
\end{equation}

Using eq. (\ref{a3}),  the quantity $g^2 A_\mu^i A_\mu ^i$
can be written in the form $g^2A_\mu^i A_\mu ^i= 4 \partial_\mu {\bf
n}\cdot \partial_\mu {\bf n}$.
Consequently, the bilinear  $ g^2 A_\mu^i A_\mu ^i$ is proportional
to the
Lagrangian of the non-linear $O(3)$ model and,  on the classical
self-dual solutions (\ref{sd}), the topological invariant $Q$ can be simply
rewritten as
\begin{equation}
\left. Q{ \over  }\right|_{\rm on\ self-dual\ sol.} = {g^2\over 32\pi}\int
d^2x\,  A_\mu^i A_\mu ^i=-{g^2\over 16\pi}\int d^2x\, {\rm Tr}[
A_\mu A_\mu ] \ \ .
\label{top3}
\end{equation}
Some remarks on eq. (\ref{top3}) are in order. First of all it is not gauge
invariant: as a matter of fact, ``small'' gauge transformations are not
allowed by the gauge fixing, while ``large'' gauge transformations change the
topological number of the vacua. Secondly,
eq. (\ref{top3}) cannot be taken as a
definition of the topological invariant $Q$, as  eq. (\ref{top3}) explicitly
depends on the metric. However,
one should remember that expression (\ref{top3}) holds only when the
self-duality condition (\ref{sd}) has been taken into account, and therefore
its validity is only ``on shell''.
 It is intriguing that, ``on shell'', the winding number $Q$ is just a mass
term for the gauge fields.

For a given $Q$, several choices of $\omega$ are possible. In turn
any choice of $\omega$ determines $n_0$, $n^a$ through (\ref{n3}).
The simplest choice of meromorphic function is
clearly a zero  of order $k$, so that one can choose $\omega (z)
=(z/z_0)^k$ where the constant $z_0 \ne 0$ has been introduced to make $\omega$
dimensionless, as required.  It is not surprising that, for such a
choice of $\omega$, the topological number evaluated through (\ref{top}) just
gives the integer power $k$. Therefore, for a given $k$, one can always
choose a pure power as the representative element of the homotopy class.
This choice of $\omega$ characterizes the radially symmetric
gauge group elements. Using eqs. (\ref{n3}) and (\ref{u3}), we obtain:

\begin{equation}
U_k= { r^{2k} - r_0^{2k}\over
r^{2k} + r_0^{2k}} -i\sigma^a E^a_r {2r_0^kr^k\over r_0^{2k} + r^{2k}}
\equiv \cos\psi - i \sigma^a E^a_r \sin\psi \ \ ,
\label{u2}
\end{equation}
where $r_0=|z_0|\ne 0$ and
$E^a_r$ is the radial component of
the following
$k-zweibein$ in polar coordinates
\begin{equation}
E^a_r=(\cos k\theta , \sin k\theta)\ , \qquad E^a_\theta=(-\sin k\theta , \cos
k\theta)\ \ ,
\label{k2}
\end{equation}
which are nothing but the usual $zweibein $ with angle
$\theta$ replaced by $k \theta$.

It is easy to check that all the basic features of the usual  polar $zweibein$
are satisfied, {\it i.e.}  they are orthonormal and cyclic  under the external
product: $E^a_\alpha E^a_\beta = \delta_{\alpha \beta}$,
 $E^a_\alpha E^b_\alpha = \delta^{ab}$,
$\epsilon^{ab}\epsilon_{\alpha \beta}E^b_\beta = E^a_\alpha$, where $\alpha ,
\beta = r, \theta$ and $\epsilon^{12}=\epsilon_{r \theta} =1$.

By construction, when $\theta$ varies from $0$ to $2 \pi$, $E^a_r$ spans
$k$ times the unit circle in the Euclidean plane. Consequently, the only
difference
with the standard polar $zweibein$ is in the derivatives of such vectors, which
provide an extra $k$ factor, {\it i.e.}
\begin{equation}
\partial_\mu E^a_\alpha= {k\over r}\epsilon_{\alpha \beta}  E^a_\beta\
e^\mu_\theta\   , \label{derivative}
\end{equation}
where we denoted by ($e^\mu_r, \ e^\mu_\theta$) the standard $zweibein$ in
polar coordinates or, equivalently, the ones given in (\ref{k2}) with $k=1$.

$U_k$ cannot be continuously deformed to the constant solution (trivial
vacuum).
Equation (\ref{u2}) with  $k=1$ is identical to the gauge group element
discussed by Jackiw and  Rebbi in ref. \cite{jac2}. For $k>1$, $U_k$  is
not equivalent to the $k$-th power of $U_1$; this power leads to a gauge
potential that does not obey the gauge condition $\partial_{\mu} A_{\mu}=0$.

The pure gauge potential corresponding to (\ref{u2}) is given by

\begin{eqnarray}
gA_\mu^3 &=&-2 {k\over r} \sin^2 \psi(r) \ e^\mu_\theta
=-{8k\over r}{r_0^{2k}r^{2k}\over (r_0^{2k} + r^{2k})^2}\ e^\mu_\theta
 \ ,\nonumber\\
gA_\mu^a E^a_r &=&2 \psi' (r)\  e^\mu_r
=-{4k\over r}{r_0^{k}r^{k}\over r_0^{2k} + r^{2k}}\ e^\mu_r \ ,\nonumber\\
gA_\mu^a E^a_\theta &=&{k\over r} \sin 2\psi (r)\  e^\mu_\theta=
{4k\over r}{r_0^{k}r^{k}(r^{2k} - r_0^{2k})\over (r_0^{2k} + r^{2k})^2}\
e^\mu_\theta \ .
\label{a}
\end{eqnarray}
All the solutions (for any $k$) are regular.

Alternatively, one could have searched for topologically non-trivial
vacua in the Abelian subgroup of $k$ radially symmetric elements of
the type
\begin{equation}
U_k= \cos\chi(r) - i \sigma^a E^a_r \sin\chi(r)\ \ .
\label{u4}
\end{equation}
The gauge condition $\partial_{\mu}A_{\mu}=0$ leads to the equation
\begin{equation}
\Delta(2 \chi) = {k^2\over r^2} \sin 2\chi\ \ .
\label{sg}
\end{equation}
In turn eq. (\ref{sg}) can be  simplified by
introducing  the variables $\varphi= 2\chi$ and  $\tau = k \log
(r/r_0)$. Thus, in terms of $\varphi$
and $\tau$, eq. (\ref{sg}) simply becomes  the one-dimensional sine-Gordon
equation
\begin{equation}
{d^2\varphi\over d\tau^2} = \sin\varphi\ ,
\label{sg2}
\end{equation}
 describing the
motion of a particle in a periodic potential $V(\varphi) =(1+\cos \varphi)$
with respect to the ``time''  $\tau =k \log (r/r_0)$.

The radially symmetric solution we have considered in eq. (\ref{u2}),
\begin{equation}
\chi(r)= \psi(r) = 2 \tan^{-1} \left( r_0\over r\right)^{k}\ \ ,
\label{psi}
\end{equation}
corresponds to
selecting the initial conditions of the particle in the periodic
potential in such a way that the
dynamics of the system is
that of a kink: the particle starts
at the ``time'' $\tau=-\infty$ from one top of the
periodic potential  and reaches at $\tau=+\infty$ a contiguous top.

Another radially symmetric solution is obtained by changing the sign of $k$
in the exponent of eq. (\ref{psi}). This change is equivalent to a conformal
transformation  and maps a kink into an anti-kink.
In spite of the fact that  our
original system ($SU(2)$ Yang--Mills in two Euclidean dimensions) is not a
conformal model, the solutions in the class we considered are
conformally invariant. Vacuum solutions are
pure gauges and therefore satisfy $F_{\mu \nu}=0$. Consequently,
the energy--momentum tensor, in particular its trace,
identically vanishes.

Starting from Eqs. (\ref{a}) one can also obtain  $2+1$ dimensional vacuum
solutions simply by replacing $r_0$ by an arbitrary function of time, {\it i.e.}
$r_0=r_0(t)$; the corresponding $gA_0 = U^{-1} \partial_0 U$ component of the
gauge potential can be easily derived from eq. (\ref{u2}). The arbitrary
reparametrization of time that such solutions exhibit is not surprising. In
fact vacuum solutions in $2+1$ dimensions follow from a
Chern--Simons action, which is invariant under diffeomorphisms.
Nevertheless the pure gauges in eq. (\ref{a}), when interpreted as $2+1$
dimensional solutions, exhibit reparametrization invariance only under the
time component: this happens because the Coulomb gauge choice breaks
diffeomorphism
invariance, leaving as a residual subgroup precisely the group of time
reparametrization.

Several arguments deserve consideration for future investigations.

At the classical level, one might wonder whether the interaction with
matter
could justify  and stabilize the particular direction in the internal
space to which
these vacuum solutions belong.
To this purpose, it seems that the most
natural choices would be an interaction either with a non-linear sigma model
or with a matter field in the adjoint representation; in this way the
direction in the internal space of the gauge group element  could be the
one induced by matter. In this framework, the method developed in refs.
\cite{nar} to find classical conformal solutions of
two-dimensional gauge theories
interacting with matter fields seems to be the most appropriate one.
Would it  be possible to find
such a matter -- Yang-Mills coupled system, in which a particular direction
in the internal space is singled out, its vacuum structure in the
vectorial sector would be identical to the
solutions of the non linear $O(3)$ model, whose quantization is well known.
Thus, the quantization of the vectorial sector would be greatly simplified;
however the possible appearance of $\theta$ vacua in the complete system 
would depend on the dynamical properties induced by matter.

The classical coincidences we have discussed could also 
be relevant in the study of lower  dimensional $QCD$ effective theories. 
To this regard, it is intriguing that the expression for the
topological invariant $Q$ 
on self-dual
solutions we have exhibited in  eq. (\ref{top3}), is identical to the leading
term of the Skyrmion Lagrangian.

\bigskip

\noindent
\acknowledgements
\noindent
We thank Roman Jackiw for useful suggestions.

\vskip 1.0truecm

{\it NOTE ADDED IN PROOF}

\vskip 1.0truecm

After the acceptance of this paper, we have been informed by
A. Jevicki that there is a sizeable, although not complete, overlap
with an investigation by A. Jevicki and N. Papanicolaou,
published in Phys. Lett. \underbar{78B}, (1978) 438.

We thank Dr. Jevicki for calling our attention on his paper.
\vfill
\eject

\end{document}